\documentclass[prd,nofootinbib,nobibnotes,twocolumn]{revtex4}
\usepackage{epsf,epsfig}
\usepackage{amssymb,amsmath,amsfonts}

\newcommand{\be}{\begin{equation}}
\newcommand{\ee}{\end{equation}}
\newcommand{\beq}{\begin{equation}}
\newcommand{\eeq}{\end{equation}}
\newcommand{\ben}{\begin{displaymath}}
\newcommand{\een}{\end{displaymath}}
\newcommand{\beqa}{\begin{eqnarray}}
\newcommand{\eeqa}{\end{eqnarray}}
\newcommand{\bea}{\begin{eqnarray}}
\newcommand{\eea}{\end{eqnarray}}
\newcommand{\bean}{\begin{eqnarray*}}
\newcommand{\eean}{\end{eqnarray*}}
\newcommand{\ba}{\begin{array}}
\newcommand{\ea}{\end{array}}
\newcommand{\bi}{\begin{itemize}}
\newcommand{\ei}{\end{itemize}}

{}{}
{}{}
\def\vereq#1#2{\lower3pt\vbox{\baselineskip1.5pt \lineskip1.5pt
\ialign{$\m@th#1\hfill##\hfil$\crcr#2\crcr\sim\crcr}}}
\makeatother

\def\be{\begin{equation}}
\def\ee{\end{equation}}
\def\bea{\begin{eqnarray}}
\def\eea{\end{eqnarray}}

\allowdisplaybreaks

\usepackage[breaklinks=true]{hyperref}
\hypersetup{   colorlinks=true,  citecolor=blue, linkcolor=red, urlcolor=blue}

\begin{document}


\title{Non-Relativistic Chern-Simons Theories \\ and Three-Dimensional Ho\v{r}ava-Lifshitz Gravity} 
        
\author{Jelle Hartong$^{a}$, Yang Lei$^{b}$,  Niels A.\ Obers$^{c}$} 
\address{\vspace{0.2cm} 
$^a$\, 
Physique Th\'eorique et Math\'ematique and International Solvay Institutes, 
Universit\'e Libre de Bruxelles, C.P. 231, 1050 Brussels, Belgium
\\
$^b$\,
Centre for Particle Theory, Department of Mathematical Sciences, 
Durham University,  South Road, Durham DH1 3LE, United Kingdom  \\
$^c$\,The Niels Bohr Institute, Copenhagen University, Blegdamsvej 17, 2100 Copenhagen \O, Denmark \\
{\small \tt hartong@ulb.ac.be, yang.lei@durham.ac.uk, obers@nbi.ku.dk}  \\
}
\begin{abstract}

We show that certain three-dimensional Ho\v{r}ava-Lifshitz gravity theories can be written as Chern-Simons gauge theories on various non-relativistic algebras. The algebras are specific extensions of the Bargmann, Newton--Hooke and Schr\"odinger algebra each of which has the Galilean algebra as a subalgebra. To show this we employ the fact that Ho\v{r}ava-Lifshitz gravity corresponds to dynamical Newton--Cartan geometry. In particular, the extended Bargmann (Newton--Hooke) Chern--Simons theory corresponds to projectable Ho\v{r}ava-Lifshitz gravity with a local $U(1)$ gauge symmetry without (with) a cosmological constant. Moreover we identify an extended Schr\"odinger algebra containing 3 extra generators that are central with respect to the subalgebra of Galilean boosts, momenta and rotations, for which the Chern--Simons theory gives rise to a novel version of non-projectable conformal Ho\v{r}ava-Lifshitz gravity that we refer to as Schr\"odinger gravity. This theory has a $z=2$ Lifshitz geometry as a vacuum solution and thus provides a new framework to study Lifshitz holography.

\end{abstract}
\maketitle



\section{Introduction} 

The local equivalence of three-dimensional Einstein gravity (with or without a cosmological constant) in terms of a Chern-Simons gauge theory \cite{Witten:1988hc,Achucarro:1987vz} has been of crucial importance in order to gain insights into the classical and quantum properties of the theory, along with holographic dualities to two-dimensional CFTs. Three-dimensional (relativistic) gravity thus plays a special role due to its simplicity, having no propagating degrees of freedom, yet being non-trivial enough to allow for black holes and numerous other interesting features. 

Recently non-relativistic geometry 
has gained considerable interest, in part due to their appearance in non-AdS holography \cite{Christensen:2013lma,Christensen:2013rfa,Hartong:2014oma,Hartong:2015wxa}, their relevance in condensed matter
setups such as the fractional quantum Hall effect \cite{Son:2013rqa,Geracie:2014nka} and other fluid/field-theoretic applications \cite{Jensen:2014aia,Hartong:2014pma,Jensen:2014ama,Hartong:2015wxa,Banerjee:2015hra}.  Moreover these geometries
lead to interesting theories of non-relativistic gravity, beyond Newtonian gravity as embodied in the original formulation of Cartan. 
In particular, a novel generalization of Newton-Cartan geometry with torsion
was first observed in \cite{Christensen:2013lma} and it was subsequently shown in \cite{Hartong:2015zia} that 
making this geometry dynamical leads to the known versions of Ho\v{r}ava-Lifshitz gravity constructed in \cite{Horava:2008ih,Blas:2009qj,Horava:2010zj}. Interesting supersymmetric extensions  of Newton-Cartan gravity have been considered as well \cite{Andringa:2013zja,Bergshoeff:2015uaa,Bergshoeff:2015ija}. 
 All this begs the question whether in three dimensions such non-relativistic gravity theories are related to Chern-Simons (CS) theories, in parallel to the relativistic case. 
 
The generalization of the CS formulation to non-relativistic Galilean gravity was initiated in the pioneering work \cite{Papageorgiou:2009zc}, 
in which the CS gauge field takes value in a Galilean algebra with two central extensions (the extended Bargmann algebra), replacing the Poincar\'e algebra 
of the relativistic setting.  We will show in this paper that this vielbein formulation is equivalent to three-dimensional torsionless Newton-Cartan (NC) gravity \cite{Hartong:2015zia}, which in turn is the 3-dimensional $U(1)$-invariant  projectable Ho\v rava--Lifshitz gravity of \cite{Horava:2010zj}.%
\footnote{The topological nature of this theory was also discussed in \cite{Horava:2010zj}.}
By going to an extended Newton-Hooke algebra, we furthermore show that a cosmological constant can be added to the theory. Moreover, by constructing a $z=2$ Schr\"odinger algebra with 3 extra generators, that are central with respect to the subalgebra of Galilean boosts, momenta and rotations, we obtain a novel action for conformal non-projectable Ho\v rava--Lifshitz gravity. The latter theory corresponds to a new version of dynamical twistless torsional Newton-Cartan geometry which we call Schr\"odinger gravity. 

The CS formulation based on the extended Bargmann algebra can be viewed as the non-relativistic counterpart of 3-dimensional Einstein gravity without a cosmological constant. Adding a cosmological constant via the Newton--Hooke algebra does not have the same effect as in the relativistic case. In particular the theory is still described by projectable Ho\v{r}ava--Lifshitz (HL) gravity. It will be shown that the cosmological constant leads to time dependent geometries. 

In order to find the counterpart of AdS$_3$ gravity we need to find a CS theory that is equivalent to non-projectable HL gravity. This is provided by considering the extended Schr\"odinger algebra in 2+1 dimensions that allows for a CS theory corresponding to twistless torsional Newton-Cartan (TTNC) gravity, or what is the same non-projectable HL gravity, with $z=2$ scaling symmetry. %
 We show that this theory of Schr\"odinger gravity admits $z=2$ Lifshitz geometries and thus provides a new framework to study Lifshitz holography. 

This letter is organized as follows. In section \ref{sec:Lie} we discuss the basic properties of the three Lie algebras on which the CS actions are based, namely the extensions of the Bargmann, Newton--Hooke and Schr\"odinger algebras that admit a non-degenerate metric. In section \ref{sec:CSactions} we construct the most general CS actions compatible with these symmetries. This includes terms that are the non-relativistic counterpart of the Lorentz CS term that can be added to the Einstein--Hilbert action in 3 dimensions. We continue in section \ref{sec:projHL} to rewrite the CS actions based on the Bargmann and Newton--Hooke algebras in the metric formulation of Newton--Cartan geometry showing that the resulting theory is a known version of projectable HL gravity. In this section we also discuss the local properties of the solutions to the flatness conditions. Finally in section \ref{sec:nonprojHL} we show that the CS theory based on the extended Schr\"odinger algebra is equivalent to a novel version of TTNC/ non-projectable HL gravity. In that section we also show that the theory admits $z=2$ Lifshitz solutions. We conclude with a discussion and outlook in section \ref{sec:discussion}.

\section{Non-relativistic Lie algebras with non-degenerate metrics}\label{sec:Lie}

Non-relativistic symmetry algebras are typically non-semisimple Lie algebras, containing the Galilean algebra as a subalgebra,
which consists (in 2+1 dimensions)  of  the generators $J$ (rotation), $P_a$ (translations, $a=1,2$), $G_a$ (Galilean boosts) and $H$  (Hamiltonian).  
In order to write down a Chern--Simons theory one needs a non-degenerate symmetric bilinear form (metric) on the Lie algebra that serves to define the trace in the Chern--Simons action%
\footnote{For brevity, the overall multiplicative constant $k/(4 \pi)$ involving the Chern--Simons level $k$,  appearing in this action
will be omitted, as  it plays no role in our discussions below.} 
\begin{equation}\label{eq:CSaction}
\mathcal{L}_{\text{CS}}=\text{Tr}\left(A\wedge dA+\frac{2}{3}A\wedge A\wedge A\right)\,.
\end{equation}
For a non-semisimple Lie algebra the existence of such a bilinear form is a non-trivial requirement, and in case of the Galilean algebra
with non-zero commutators 
\begin{equation}\label{eq:Gal}
[J\,,P_a]=\epsilon_{ab}P_b \,,\ \  [J\,,G_a]=\epsilon_{ab}G_b\,,\ \  [H\,,G_a]=P_a\,,
\end{equation}
it necessitates the addition of central elements. While in any dimension the Galilean algebra can be centrally extended to the Bargmann algebra
using the mass
generator $N$ satisfying 
\begin{equation}\label{eq:Barg}
[P_a\,,G_b]=N\delta_{ab}\,,
\end{equation}
in 2+1 space-time it is possible to add three further central elements $S$, $Y$ and $Z$  as follows\footnote{In the case of the Galilei algebra one cannot add the $Y$ generator as a central extension because there is no non-trivial cohomology associated with it as follows from the results of \cite{Bonanos:2008kr,Alvarez:2007fw}. Here we will never use $Y$ in the context of the Galilei algebra but only in the larger Schr\"odinger algebra. We thank Joaquim Gomis for pointing this out to us.}
\begin{eqnarray}\nonumber
&& [G_a\,,G_b]=S \epsilon_{ab}\,,\quad [P_a\,, P_b] =Z \epsilon_{ab}  \,, \\ \label{eq:centralext} &&  [P_a\,,G_b] =N\delta_{ab} -Y\epsilon_{ab}\,.
\end{eqnarray}
These play an important role in obtaining non-degenerate metrics on various non-relativistic symmetry algebras such as the Bargmann, Newton--Hooke and Schr\"odinger algebras. In the following, we denote by $B(x,y)$ the bilinear form where $x$ and $y$ are elements of the Lie algebra. Symmetry requires that $B(x,y)=B(y,x)$ and invariance under the action of the algebra corresponds to $B([z,x],y)+B(x,[z,y])=0$ for all $z,x,y$.

\subsection{Extended Bargmann algebra}\label{subsec:extBarg}

If we add the central element $S$ in  \eqref{eq:centralext} (but not $Y$ and $Z$) to the Bargmann algebra \eqref{eq:Gal}, \eqref{eq:Barg} the resulting non-semisimple Lie algebra is a semi-direct sum of the normal subalgebra $H,P_a,N$ with the Nappi--Witten algebra \cite{Nappi:1993ie} consisting of $J, G_a, S$  (which is a central extension of the 2-dimensional Euclidean algebra). 
This algebra was used in the Chern--Simons theory (CS) of \cite{Papageorgiou:2009zc} and corresponds, as shown below, to a 3D projectable Ho\v rava--Lifshitz gravity theory.  The possible non-trivial values of $B(x,y)$ for the centrally extended Bargmann algebra are given by
\begin{eqnarray}\nonumber
&& B(H,S)=-B(J,N)=c_1\,,\quad B(P_a, G_b)= c_1 \epsilon_{ab}\,,  \\ \nonumber
&& B(G_a, G_b) =c_2 \delta_{ab}\,, \quad B(J,S)=c_2\,,\\ \label{bargmanbili}
&& B(J,J)=c_3\,, \quad B(H,J)=c_4\,, \quad B(H,H)=c_5\,,
\end{eqnarray}
with $c_i$ arbitrary constants and with $c_1\neq 0$ for the matrix to be non-degenerate. If we remove the central element $S$ from the algebra the bilinear form becomes degenerate.

\subsection{Extended Newton--Hooke algebra}

There exists a deformation of the Bargmann algebra called the Newton--Hooke algebra. Its nonzero commutators are those of  \eqref{eq:Gal}, \eqref{eq:Barg} plus $[H\,,P_a]=-\Lambda_c G_a\,$.
There exists an extension of this algebra involving the $S$ generator where the central element appears in
\begin{eqnarray}
&&\left[G_a\,, G_b\right]=S\epsilon_{ab}\,,\ \ \left[H\,,P_a\right]=-\Lambda_c G_a\,,\nonumber\\ 
&&\left[P_a\,,P_b\right]=\Lambda_c S\epsilon_{ab}\,.\label{eq:extendedNH}
\end{eqnarray}
This extended Newton--Hooke algebra, which reduces to the extended Bargmann algebra for $\Lambda_c=0$, 
 was studied in the context of CS theories in \cite{Papageorgiou:2010ud}.  For  $\Lambda_c\neq 0$, the parameter $\Lambda_c$ can be set to one by rescaling $(H,P_a,N) \rightarrow\Lambda_c^{1/2}  (H,P_a,N) $. 
The most general symmetric bilinear form that one can define on the algebra is given by \eqref{bargmanbili} together with
\begin{eqnarray}\label{eq:Bilinear}
B(H,N)=-\Lambda_c c_2\,, \qquad B(P_a, P_b)= \Lambda_c c_2 \delta_{ab}\,,
\end{eqnarray}
and requiring $\Lambda_c\neq c_1^2/c_2^2$ ensures that the matrix is non-degenerate.

\subsection{Extended Schr\"odinger algebra}

The conformal extension of the Bargmann algebra is the Schr\"odinger algebra (with dynamical exponent $z=2$). The Hamiltonian is extended to an $SL(2,\mathbb{R})$ algebra consisting of dilatations $D$ with $z=2$ and a special conformal generator $K$ that form the subalgebra
\begin{equation}\label{eq:SL2R}
[D\,,H]=-2H\,, \quad [H\,,K]=D\,, \quad [D\,,K]=2K\,.
\end{equation}
The Schr\"odinger algebra is obtained by taking this $SL(2,\mathbb{R})$ algebra and specifying how it acts on the Bargmann subalgebra \eqref{eq:Gal}, \eqref{eq:Barg}. This action is given by
\begin{eqnarray}\nonumber
&& [H\,,G_a]=P_a\,,\quad [D\,,P_a]= -P_a\,,\\ \label{eq:SL2RonSch}&&  [D\,,G_a] =G_a\,,\quad [K\,,P_a]=-G_a\,.
\end{eqnarray}
The mass generator $N$ remains central with respect to the full Schr\"odinger algebra. 

It is possible to add dilatations to the extended Bargmann algebra of section \ref{subsec:extBarg} by taking $[D,S]=2S$. However this algebra has no non-degenerate metric. If we consider the full central extension \eqref{eq:centralext}, i.e. we add  $S$, $Y$ and $Z$ to the Bargmann algebra we can add the full $SL(2,\mathbb{R})$ algebra \eqref{eq:SL2R} such that \eqref{eq:SL2RonSch} continue to hold. The action of the $SL(2,\mathbb{R})$ subalgebra on $S$, $Y$ and $Z$ is non-trivial and fully determined by the Jacobi identities given all the other commutators.%
\footnote{This is an explicit example of a more general theorem on double extensions of Lie algebras, elaborated on in \cite{FigueroaO'Farrill:1995cy}. We thank Jan Rosseel for useful discussions on this point.} 
 The result is that the nonzero commutators are
\begin{eqnarray}
&&\hspace{-1cm} [H\,,Y]=-Z\,, \quad [H\,,S] =-2Y\,,\quad [K\,,Y]=S\,, \nonumber\\
&&\hspace{-1cm} [K\,,Z]=2Y\,, \quad [D\,,S]=2S\,,\quad [D\,,Z]=-2Z\,.  \label{eq:SL2Roncentral}
\end{eqnarray}
The extended Schr\"odinger algebra is thus given by \eqref{eq:Gal}--\eqref{eq:centralext}, \eqref{eq:SL2R}, \eqref{eq:SL2RonSch} and \eqref{eq:SL2Roncentral}. The corresponding symmetric bilinear form invariant under the extended Schr\"odinger algebra is 
\begin{eqnarray}\nonumber
&& B(H,S)=B(D,Y)=B(K,Z)=-B(J,N)=c_1\,,\\ \nonumber &&  B(P_a,G_b)= c_1 \epsilon_{ab} \quad B(H,K)=-c_2\,,\\ \label{eq:BilinearSch}
&&  B(D,D)=2c_2\,, \quad B(J,J)=c_3\,,
\end{eqnarray}
which is non-degenerate if $c_1 \neq 0$.

\section{Non-relativistic Chern--Simons actions}\label{sec:CSactions}

We now turn to study the form of the CS action \eqref{eq:CSaction}  for each of these  three algebras which have the Bargmann algebra as a subalgebra and allow for a non-degenerate metric. 

\subsection{Bargmann and Newton--Hooke invariant Chern--Simons actions}\label{subsec:CSBarg}

The extended Bargmann algebra can be obtained by setting $\Lambda_c=0$ in the extended Newton--Hooke algebra so we will construct the CS action using the metric \eqref{bargmanbili} and \eqref{eq:Bilinear}.
Expanding the gauge connection as
$
A=H\tau+P_ae^a+G_a\Omega^a+J\Omega+Nm+S\zeta\,,
$
the CS action becomes 
\begin{eqnarray}\nonumber
&&\text{Tr}\left(A\wedge dA+\frac{2}{3}A\wedge A\wedge A\right)=\\ \nonumber
&&2c_1\Big[-\epsilon_{ab} R^a(G)\wedge e^b+\frac{1}{2}\epsilon_{ab}\tau\wedge\Omega^a\wedge\Omega^b-\Omega\wedge dm\\ \nonumber &&+ \zeta\wedge d\tau+  \Lambda_c\tau\wedge e^1\wedge e^2\Big]\\ \nonumber
&&+c_2\big[\Omega^a\wedge R^a(G)+2\zeta\wedge d\Omega+\Lambda_c e^a\wedge R^a(P) \\ \nonumber && -2\Lambda_c\tau\wedge R(N)+\Lambda_c e^a\wedge\Omega^a\wedge\tau\big]\\ \label{eq:CStheory}
&&+c_3\Omega\wedge d\Omega+2c_4\tau\wedge d\Omega+c_5\tau\wedge d\tau\,,
\end{eqnarray}
(see also \cite{Papageorgiou:2009zc,Papageorgiou:2010ud}) where the curvatures $R^a(P)$, $R^a(G)$ and $R(N)$ are given by
\begin{eqnarray}\label{eq:curv}
R^a(P) &=& de^a-\Omega^a\wedge\tau-\epsilon^{ab}\Omega\wedge e^b\,,\\ \nonumber R^a(G) &=& d\Omega^a-\epsilon^{ab}\Omega\wedge\Omega^b\,,\quad R(N) = dm-\Omega^a\wedge e^a\,.
\end{eqnarray}
These curvatures are defined by the expansion of the field strength
\begin{eqnarray}\nonumber
F &=& dA+A\wedge A \\ \nonumber &=& HR(H)+P_aR^a(P)+G_aR^a(G)+JR(J) \\ \label{eq:expF} &&+NR(N)+SR(S)\,.
\end{eqnarray}
We see that $\Lambda_c$ plays the role of a cosmological constant term (in the $c_1$ term). The terms proportional to $c_2\Lambda_c$ are by themselves invariant under the gauge transformations $\delta A=d\Lambda+[A\,,\Lambda]$.

The terms with coefficients $c_4$ and $c_5$ in \eqref{eq:CStheory} are not interesting as they can be removed by a field redefinition of $\zeta$. This leads to a new value for the parameter in front of the $\Omega\wedge d\Omega$ term. 
Hence we can always restrict ourselves to $c_1$, $c_2$ and $c_3$ and set to zero $c_4=c_5=0$. When $\Lambda_c=0$ the terms proportional to $c_2$ and $c_3$ are
\begin{equation}\label{eq:GboostCS}
c_2\left(\Omega^a\wedge R^a(G)+2\zeta\wedge d\Omega\right)+c_3\Omega\wedge d\Omega\,.
\end{equation}
These can be thought of as the analogue of the Lorentz CS term. The term with coefficient $c_2$ is a novel Galilean boost invariant combination that starts as $\Omega^a\wedge d\Omega^a$ plus extra terms to make it invariant. To see the invariance explicitly we give the transformations of the connections for $\Lambda_c=0$ appearing in \eqref{eq:GboostCS} that read
\begin{eqnarray}
\delta\Omega^a & = & d\lambda^a+\epsilon^{ab}\left(\lambda\Omega^b-\lambda^b\Omega\right)\,,\qquad\delta\Omega = d\lambda\,,\nonumber\\
\delta\zeta & = & -\epsilon^{ab}\lambda^a\Omega^b\,.
\end{eqnarray}
If we consider the CS theory on a manifold with a boundary  they are expected to lead to Galilean boost and rotation anomalies on the boundary theory. In the simplest setting with $c_2=c_3=0$ the $\zeta$ equation of motion is $d\tau=0$. In section \ref{sec:HLactions} we will see that this corresponds to having no torsion in the Newton--Cartan description, or what is the same, projectable HL gravity \cite{Hartong:2015zia}.

\subsection{Schr\"odinger invariant Chern--Simons action \label{subsec:CSSch}}

The extended Schr\"odinger algebra is \eqref{eq:Gal}--\eqref{eq:centralext}, \eqref{eq:SL2R}, \eqref{eq:SL2RonSch} and \eqref{eq:SL2Roncentral}. We expand the gauge field as 
\begin{eqnarray}
A & = & H \tau +P_a e^a +G_a \omega^a + J \omega + N m +D b +K f \nonumber\\
&&+S\zeta+Y \alpha  + Z\beta \,.\label{eq:ASch}
\end{eqnarray}
Using the metric on the Lie algebra \eqref{eq:BilinearSch} the Chern--Simons action can be written as
\begin{eqnarray}
\mathcal{L} &=& 2c_1\Big[\tilde R^2(G) \wedge e^1 -\tilde R^1(G)\wedge e^2 +\tau \wedge \omega^1 \wedge \omega^2 \nonumber \\ && -m\wedge d\omega - f \wedge e^1 \wedge e^2 +\zeta \wedge (d\tau-2b\wedge \tau) \nonumber \\ 
 && +\alpha \wedge (db -f \wedge \tau) +\beta \wedge (df+ 2b\wedge f)\Big] \label{frlag}\\
 &&+2c_2\left[b\wedge db-\tau\wedge df+2b\wedge\tau\wedge f\right]+c_3\omega\wedge d\omega\,,\nonumber
\end{eqnarray}
where the curvature $\tilde R^a(G)$ is given by
\begin{equation}
\tilde R^a(G) = d\omega^a +\epsilon^{ab}\omega^b \wedge \omega -\omega^a \wedge b-f \wedge e^a\,.
\end{equation}

There is no redefinition of the connections $\zeta$, $\alpha$ and $\beta$ that allows one to remove the term with coefficient $c_2$ entirely. It transforms under the $SL(2,\mathbb{R})$ transformations inside the extended Schr\"odinger algebra. It would be interesting to see if it corresponds to some anomaly for a boundary theory like a Weyl-type anomaly.

The equation of motion of $\zeta$ now imposes the on-shell condition that $d\tau = 2b \wedge \tau$ which is equivalent to $\tau\wedge d\tau=0$. In the language of Newton--Cartan geometry this corresponds to twistless torsional Newton--Cartan (TTNC) geometry \cite{Christensen:2013lma,Bergshoeff:2014uea} or what is the same non-projectable HL gravity \cite{Hartong:2015zia}. The details will be given in section \ref{sec:HLactionsSch}.

\section{Chern--Simons action for 3D projectable Ho\v rava--Lifshitz gravity}\label{sec:projHL}

We know from \cite{Andringa:2010it} that gauging the Bargmann algebra leads to Newton--Cartan (NC) geometry. In \cite{Hartong:2015zia} it was shown that dynamical Newton--Cartan geometry is field redefinition equivalent to projectable Ho\v rava--Lifshitz gravity as presented in \cite{Horava:2010zj}. Hence we should be able to show that the CS action given in  section \ref{subsec:CSBarg} is equivalent to a 3D projectable HL gravity theory.

\subsection{Bargmann invariant projectable Ho\v rava--Lifshitz gravity}\label{sec:HLactions}

We will now rewrite \eqref{eq:CStheory} with only the $c_1$ coefficient nonzero in a metric form using the language of Newton--Cartan (NC) geometry. The connections $\tau_\mu$ and $e^a_\mu$ are the vielbeins of NC geometry. We define inverse vielbeins $v^\mu$ and $e^\mu_a$ via $\delta^\mu_\nu=-v^\mu\tau_\nu+e^\mu_a e^a_\nu$ so $v^\mu \tau_\mu=-1$, $e^\mu_a\tau_\mu=0$, $v^\mu e^a_\mu=0$ and $e_a^\mu e_\mu^b=\delta_a^b$. 
It can be shown that the first term in the CS action \eqref{eq:CStheory} can be written as
\begin{equation}
R^2(G)\wedge e^1-R^1(G)\wedge e^2=v^\mu e^\nu_a R_{\mu\nu}{}^a(G)\tau\wedge e^1\wedge e^2\,.
\end{equation}
With $m=-v^\mu m_\mu\tau+e^\mu_a m_\mu e^a$ it follows that the third term in \eqref{eq:CStheory} becomes
\begin{eqnarray}\nonumber
m\wedge R(J)=\Big(-\frac{1}{2}v^\mu m_\mu\mathcal{R}-e^\rho_2 m_\rho v^\mu e^\nu_1 R_{\mu\nu}(J) \\ +e^\rho_1 m_\rho v^\mu e^\nu_2 R_{\mu\nu}(J)\Big)\tau\wedge e^1\wedge e^2\,,
\end{eqnarray}
where we used that 
\begin{equation}\label{eq:curlR}
R_{ab}(J)=e^\mu_a e^\nu_b R_{\mu\nu}(J)\equiv\frac{1}{2}\epsilon_{ab}\mathcal{R}\,.
\end{equation}

The action \eqref{eq:CStheory} is written in a first order formalism where all the connections in $A_\mu$ are treated as independent variables. The form we are looking for treats the NC variables $\tau_\mu$, $e^a_\mu$ and $m_\mu$ as the independent variables. Hence we will integrate out the variables $\Omega^a$, $\Omega$ and $\zeta$. Their equations of motion are the NC curvature constraints \cite{Andringa:2010it} $R^a(P)=0$, $R(N)=0$ and $R(H)=d\tau=0$ where the curvatures are given in \eqref{eq:curv}. These are solved by expressing $\Omega^a_\mu$ and $\Omega_\mu$ in terms of $\tau_\mu$, $e^a_\mu$ (their inverse) and $m_\mu$ where $d\tau=0$. The off-shell implementation of the curvature constraints makes the theory diffeomorphism invariant because the NC curvature constraints imply that the transformations of $\tau_\mu$, $e^a_\mu$ and $m_\mu$ constitute diffeomorphisms and local $G^a$, $J$, $N$ transformations \cite{Andringa:2010it}.

In order to rewrite the CS action it will be useful to employ the following Bianchi identity
\begin{eqnarray}
&&dR^a(P)-\epsilon^{ab}\Omega\wedge R^b(P)-\Omega^a\wedge d\tau \nonumber\\
&&=-R^a(G)\wedge\tau-\epsilon^{ab}R(J)\wedge e^b\,.
\end{eqnarray}
Using the curvature constraints $R^a(P)=0$ and $d\tau=0$ which will be implemented off-shell we find $R^a(G)\wedge\tau+\epsilon^{ab}R(J)\wedge e^b=0$. From this we conclude that
\begin{eqnarray}
&&v^\mu e^\nu_1 R_{\mu\nu}(J)=-e^\mu_2 e^\nu_a R_{\mu\nu}{}^a(G)\,,\nonumber\\
&&v^\mu e^\nu_2 R_{\mu\nu}(J)=e^\mu_1 e^\nu_a R_{\mu\nu}{}^a(G)\,.
\end{eqnarray}
Using that $\Omega^1=-v^\mu\Omega_\mu^1\tau+e^\mu_a\Omega^1_\mu e^a$ we conclude that \eqref{eq:CStheory}, with $c_1=1$ and all other constants zero, can be written as
\begin{eqnarray}
\mathcal{L} & = & e \left(2\hat v^\mu e^\nu_a R_{\mu\nu}{}^a(G)+\left(e^\mu_a e^\nu_b-e^\nu_a e^\mu_b\right)\Omega_\mu^a\Omega_\nu^b\right.\nonumber\\
&&\left.+v^\mu m_\mu\mathcal{R}\right)\,,
\end{eqnarray}
where $e=\tau\wedge e^1\wedge e^2$.

To massage this expression further we  need a notion of a covariant derivative. This can be introduced via the vielbein postulates
\begin{eqnarray}\nonumber
\mathcal{D}_\mu\tau_\nu &=&\partial_\mu\tau_\nu-\Gamma^\rho_{\mu\nu}\tau_\rho=0\,,\\  \mathcal{D}_\mu e^a_\nu &=& \partial_\mu e^a_\nu-\Gamma^\rho_{\mu\nu}e^a_\rho-\Omega^a_\mu\tau_\nu-\epsilon^{ab}\Omega_\mu e^b_\nu=0\,,
\end{eqnarray}
where we take for $\Gamma_{\mu\nu}^\rho$
\begin{equation}\label{eq:NCconnection1}
\Gamma^\rho_{\mu\nu}=-\hat v^\rho\partial_\mu\tau_\nu+\frac{1}{2}h^{\rho\sigma}\left(\partial_\mu\bar h_{\nu\sigma}+\partial_\nu\bar h_{\mu\sigma}-\partial_\sigma\bar h_{\mu\nu}\right)\,,
\end{equation}
in which
\begin{eqnarray}\nonumber 
&& \hat v^\mu=v^\mu-h^{\mu\nu}m_\nu\,,\quad \bar h_{\mu\nu}=h_{\mu\nu}-\tau_\mu m_\nu-\tau_\nu m_\mu\,,\\ \label{eq:hatvbarh} && h_{\mu\nu}=\delta_{ab}e^a_\mu e^b_\nu\,,\quad h^{\mu\nu}=\delta^{ab}e_a^\mu e_b^\nu\,.
\end{eqnarray}
The connection \eqref{eq:NCconnection1} is a symmetric connection for $d\tau=0$ that is invariant under $G^a$, $J$, and $N$ transformations. The vielbein postulates relate $\Gamma^\rho_{\mu\nu}$ to $\Omega^a_\mu$ and $\Omega_\mu$. These relations are the same as the expressions obtained by solving the curvature constraints $R_{\mu\nu}^a(P)=0$, $R_{\mu\nu}(N)=0$ for $\Omega^a_\mu$ and $\Omega_\mu$. We denote by $\nabla_\mu$ the covariant derivative containing the connection $\Gamma^\rho_{\mu\nu}$. For $d\tau=0$ we have \cite{Andringa:2010it,Hartong:2015zia}
\begin{eqnarray}\nonumber 
&& \left[\nabla_\mu\,,\nabla_\nu\right] X_\sigma = R_{\mu\nu\sigma}{}^\rho X_\rho\,,\\ \label{eq:Riem} && R_{\mu\nu\sigma}{}^\rho=e^\rho_a\tau_\sigma R_{\mu\nu}{}^a(G)-e_{\sigma a}e^\rho_b R_{\mu\nu}(J)\epsilon^{ab}\,.
\end{eqnarray}

We now switch to employing a Lagrangian density rather then a 3-form. Using \eqref{eq:Riem} and the fact that from the vielbein postulates it follows that $\Omega_\mu^a e^\nu_a=\nabla_\mu v^\nu$ we find after performing a few partial integrations and writing $v^\mu=\hat v^\mu+h^{\mu\nu}m_\nu$ that
\begin{eqnarray}\nonumber
\mathcal{L} &=& e\Big(-\nabla_\mu\hat v^\mu\nabla_\nu\hat v^\nu+\nabla_\nu\hat v^\mu\nabla_\mu\hat v^\nu+v^\mu m_\mu\mathcal{R} \\  && +(h^{\mu\rho}h^{\nu\sigma}-h^{\mu\sigma}h^{\nu\rho})\nabla_\mu m_\rho\nabla_\nu m_\sigma \Big)\,.
\end{eqnarray}
Finally, using partial integrations which give rise to a commutator on one of the $m_\mu$ vectors as well as properties of the Riemann tensor, it can be shown that
\begin{equation}\label{eq:U(1)invmodel}
\mathcal{L}=e\left(h^{\mu\rho}h^{\nu\sigma}K_{\mu\nu}K_{\rho\sigma}-\left(h^{\mu\nu}K_{\mu\nu}\right)^2-\tilde\Phi\mathcal{R}\right)\,,
\end{equation}
where 
\begin{equation}
\tilde\Phi=-v^\mu m_\mu+\frac{1}{2}h^{\mu\nu}m_\mu m_\nu\,. 
\end{equation}
This is the same action as the action\footnote{In the analysis of \cite{Hartong:2015zia} a different choice was made for the connection $\Gamma^\rho_{\mu\nu}$ that was denoted by $\hat\Gamma^\rho_{\mu\nu}$. This other choice is related to \eqref{eq:NCconnection} via equations (5.7) and (5.3) of \cite{Hartong:2015zia}. It can be shown that the form of the Lagrangian is not affected by these choices.} (10.10) given in \cite{Hartong:2015zia} which in turn is based on the NC version of the results of \cite{Horava:2010zj}. Note that the extrinsic curvature is given by $h^{\nu\rho}K_{\mu\rho}=-\nabla_\mu\hat v^\nu$. One observes that the HL $\lambda$ parameter which can appear between the two extrinsic curvature terms is equal to unity in \eqref{eq:U(1)invmodel}

If we include $\Lambda_c$ appearing in the extended Newton--Hooke algebra we simply end up with the same Lagrangian to which we add $e\Lambda_c$. We note that the sign of the cosmological constant term is not fixed.

The Lagrangian \eqref{eq:U(1)invmodel} should be thought of as depending on the variables $\tau_\mu=\partial_\mu\tau$, $\tilde\Phi$ and $\bar h_{\mu\nu}$ and their derivatives. In projectable HL gravity $\tau$ is identified with the ADM time coordinate leading to foliation preserving diffeomorphism invariance.

\subsection{Solutions}

We will solve the equations of motion of \eqref{eq:CStheory},  $F=0$ with $F$ expanded as in \eqref{eq:expF}, locally for the case with $c_2=c_3=c_4=c_5=0$ but with $\Lambda_c$ arbitrary. Under a gauge transformation the connection transforms as $\delta A=d\Lambda+[A,\Lambda]$. We will write $\Lambda$ as $\Lambda=\xi^\mu A_\mu+\Sigma\,$, where $\Sigma=G_a\lambda^a+J\lambda+N\sigma+S\kappa$. In components these are the following transformations
\begin{eqnarray}
&&\delta\tau_\mu = \mathcal{L}_\xi\tau_\mu\,,\quad\delta e^a_\mu = \mathcal{L}_\xi e^a_\mu+\lambda^a\tau_\mu+\epsilon^{ab}\lambda e^b_\mu\,,\nonumber \\ 
&&\delta\Omega^a_\mu = \mathcal{L}_\xi\Omega^a_\mu+\partial_\mu\lambda^a+\epsilon^{ab}\left(\lambda\Omega_\mu^b-\lambda^b\Omega_\mu\right)\,,\nonumber\\
&&\delta m_\mu = \mathcal{L}_\xi m_\mu+\lambda^a e^a_\mu+\partial_\mu\sigma\,,\quad\delta\Omega_\mu = \mathcal{L}_\xi\Omega_\mu+\partial_\mu\lambda\,,\nonumber \\ && \delta\zeta_\mu = \mathcal{L}_\xi\zeta_\mu-\epsilon^{ab}\lambda^a\Omega^b_\mu+\partial_\mu\kappa\,.
\end{eqnarray}

Without loss of generality we can fix the gauge redundancy by setting $\tau_\mu=\delta_\mu^t$, $e^a_\mu=\delta_\mu^i\delta_i^a$, $\Omega_\mu=0$, $\Omega^a =-\Lambda_c\delta^a_i x^i$ and $m=\frac{1}{2}\Lambda_c x^ix^idt+d\sigma$. The relation between NC geometry and the ADM form of the HL metric
\begin{equation}\label{eq:ADM}
ds^2=-N^2dt^2+\gamma_{ij}\left(dx^i+N^idt\right)\left(dx^j+N^jdt\right)\,,
\end{equation}
uses the following identifications (see section 8 of \cite{Hartong:2015zia})
\begin{eqnarray}\nonumber
&& \tau_t=N\,,\tau_i=0\,,\;\quad h_{ij}=\gamma_{ij}\,,\;h_{it}=h_{tt}=0\,,\\ \label{eq:HLmetric} && m_t=0\,, \; m_i=-N^{-1}\gamma_{ij}N^j\,.
\end{eqnarray}
This identification only works in special gauges of the CS theory.  When written in the form \eqref{eq:U(1)invmodel}
the HL theory is not a Lorentzian metric theory. In order to make contact with the ADM parametrization we take $
\sigma=-\frac{1}{2}\Lambda_c t x^ix^i\,,
$
so that $m_t=0$ and $m_i=\partial_i\sigma=-\Lambda_c t x^i\,.
$

Hence the full solution for $\tau$, $e^a$ and $m$ is given by $\tau=dt$, $e^a=\delta^a_idx^i$, $m=-\Lambda_c t x^idx^i$. This corresponds to the ADM variables
$
N=1\,,\  N^i=\Lambda_c t x^i\,,\  h_{ij}=\delta_{ij}\,.
$
By making the  coordinate transformation $x^i=e^{-\Lambda_c t^2/2}X^i$ this becomes
\begin{equation}
ds^2=-dt^2+e^{-\Lambda_c t^2}dX^i dX^i\,.
\end{equation}
We thus find cosmological solutions for $\Lambda_c\neq 0$. Of course this is only true sufficiently locally,
as there can be  non-trivial identifications on a global level.

\section{Chern--Simons actions for 3D non-projectable Ho\v rava--Lifshitz gravity}\label{sec:nonprojHL}

In \cite{Bergshoeff:2014uea} it was shown that gauging the Schr\"odinger algebra leads to torsional Newton--Cartan geometry with twistless torsion $\tau\wedge d\tau=0$. In \cite{Hartong:2015zia} it has been shown that twistless torsional Newton--Cartan geometry (TTNC) corresponds to non-projectable HL gravity. We refer to \cite{Afshar:2015aku} for an alternative derivation of the same connection between dynamical TTNC geometry and HL gravity.
We now show that the CS action given in  section \ref{subsec:CSSch} is equivalent to a 3D non-projectable HL gravity theory.

\subsection{Schr\"odinger gravity}\label{sec:HLactionsSch}

Our goal will be to rewrite the CS Lagrangian \eqref{frlag} with $c_2=c_3=0$ into the metric formulation of TTNC geometry. 
As in the case discussed in section \ref{sec:HLactions} we will go from a first order formalism to a second order one by integrating out the connections $\omega^a$, $\omega$, $\zeta$ and $\alpha$. The equations of motion corresponding to varying these connections are the curvature constraints $\tilde R^a(P)=0$, $\tilde R(N)=0$, $\tilde R(H)=0$, and $\tilde R(D)=0$. These curvatures can be computed by expanding the curvature of \eqref{eq:ASch} as
\begin{eqnarray}
F & = & H\tilde R(H)+P_a\tilde R^a(P)+G_a\tilde R^a(G)+J\tilde R(J)\nonumber  \\   &&+N\tilde R(N) +D\tilde R(D)+K\tilde R(K)+S\tilde R(S)\nonumber\\&&+Y\tilde R(Y)+Z\tilde R(Z)\,.
\end{eqnarray}
Solving the constraints $\tilde R^a(P)=0$, $\tilde R(N)=0$, $\tilde R(H)=0$ and $\tilde R(D)=0$ was done in \cite{Bergshoeff:2014uea} and the solution can be expressed as giving $\omega^a$, $\omega$, $b$ and $f$ in terms of the vielbeins $\tau$ (obeying $\tau\wedge d\tau=0$), $e^a$, $m$ and the components $\hat v^\mu b_\mu$ and $\hat v^\mu f_\mu$. The curvature constraints also allow us to rewrite the algebra of gauge transformations acting on these fields as the algebra of diffeomorphisms and internal transformations consisting of local $G^a$, $J$, $N$, $D$ and $K$ transformations. 

The expressions for $\omega^a$ and $\omega$ can also be obtained from a vielbein postulate for a specific realization of an affine connection $\tilde\Gamma^\rho_{\mu\nu}$ that is invariant under all the transformations except those that are diffeomorphisms. These vielbein postulates are
\begin{eqnarray}
\mathcal{D}_\mu\tau_\nu &=& \partial_\mu\tau_\nu-\tilde\Gamma^\rho_{\mu\nu}\tau_\rho-2b_\mu\tau_\nu=0\,,\\ \nonumber  \mathcal{D}_\mu e^a_\nu&=&\partial_\mu e^a_\nu-\tilde\Gamma^\rho_{\mu\nu}e^a_\rho-\omega^a_\mu\tau_\nu-\epsilon^{ab}\omega_\mu e^b_\nu-b_\mu e^a_\nu=0\,,
\end{eqnarray}
where we take for $\tilde\Gamma_{\mu\nu}^\rho$
\begin{eqnarray}\nonumber
\tilde\Gamma^\rho_{\mu\nu} &=&-\hat v^\rho\left(\partial_\mu-2b_\mu\right)\tau_\nu+\frac{1}{2}h^{\rho\sigma}\Big((\partial_\mu-2b_\mu)\bar h_{\nu\sigma} \\ \label{eq:NCconnection} && +(\partial_\nu-2b_\nu)\bar h_{\mu\sigma}-(\partial_\sigma-2b_\sigma)\bar h_{\mu\nu}\Big)\,.
\end{eqnarray}
The connection $\tilde\Gamma^\rho_{\mu\nu}$ is symmetric. The associated curvature is 
$
[\tilde\nabla_\mu\,,\tilde\nabla_\nu]X_\sigma=\tilde R_{\mu\nu\sigma}{}^\rho X_\rho\,
$
for any vector $X_\rho$ where \cite{Hartong:2015zia}
\begin{eqnarray}\nonumber 
\tilde R_{\mu\nu\sigma}{}^\rho&=&-e^{\rho d}e^c_\sigma\epsilon_{cd}\tilde R_{\mu\nu}(J)+e^{\rho}_c\tau_\sigma\tilde R_{\mu\nu}^c(G) \\&&-\delta^\rho_\mu\tau_\sigma f_\nu+\delta^\rho_\nu\tau_\sigma f_\mu+\delta^\rho_\sigma\left(f_\mu\tau_\nu-f_\nu\tau_\mu\right)\,. \qquad 
\end{eqnarray}
The equations of motion for $\zeta$ and $\alpha$ are solved by
\begin{eqnarray}\nonumber
b_\nu &=& \frac{1}{2}\hat v^\mu\left(\partial_\mu\tau_\nu-\partial_\nu\tau_\mu\right)-\hat v^\mu b_\mu\tau_\nu\,,\\  f_\nu &=& \hat v^\mu\left(\partial_\mu b_\nu-\partial_\nu b_\mu\right)-\hat v^\mu f_\mu\tau_\nu\,,
\end{eqnarray}
which is why we are left with $\hat v^\mu b_\mu$ and $\hat v^\mu f_\mu$ as independent variables on top of the usual TTNC variables $\tau$, $e^a$ and $m$. These expressions satisfy $e^\mu_a e^\nu_b R_{\mu\nu}(K)=0$.

Using the curvature constraints the Lagrangian \eqref{frlag} for $c_2=c_3=0$ and $c_1=1$ can be written as
\begin{eqnarray}\nonumber
\mathcal{L}&=&2\big(e^a\wedge\omega^a\wedge\omega-\tau\wedge\omega^1\wedge\omega^2+f\wedge e^1\wedge e^2 \\ &&+\beta\wedge\left(df+2b\wedge f\right)\big)\,.
\end{eqnarray}
With the help of the vielbein postulates this can be further rewritten as
\begin{eqnarray}\nonumber
\mathcal{L} &=& -\Big(2\epsilon^{\mu\nu\rho}m_\rho\partial_\mu\omega_\nu+\epsilon^{\mu\nu\rho}\epsilon_{\sigma\lambda\kappa}\tau_\rho v^\kappa\tilde\nabla_\mu v^\sigma\tilde\nabla_\nu v^\lambda \\ && +2\hat v^\mu f_\mu\Big)\tau\wedge e^1\wedge e^2+2\beta\wedge\left(df+2b\wedge f\right).
\end{eqnarray}
Using the above mentioned results multiple times as well as \eqref{eq:hatvbarh} and after performing various partial integrations a lengthy calculation gives
\begin{eqnarray}\nonumber 
\mathcal{L} &=& e\Big[\left(h^{\alpha\nu}h^{\beta\mu}-h^{\alpha\mu}h^{\beta\nu}\right)\bar h_{\alpha\sigma}\tilde\nabla_\mu\hat v^\sigma \bar h_{\beta\lambda}\tilde\nabla_\nu\hat v^\lambda-\tilde\Phi\tilde{\mathcal{R}} \\ && -2\hat v^\mu f_\mu+2\epsilon^{\mu\nu\rho}\tau_\rho\hat v^\sigma\beta_\nu R_{\mu\sigma}(K)\Big]\,,
\end{eqnarray}
where we defined
$
\tilde R_{ab}(J)=e^\mu_a e^\nu_b\tilde R_{\mu\nu}(J)\equiv\frac{1}{2}\epsilon_{ab}\tilde{\mathcal{R}}\,.
$

The next step is to go from the connection $\tilde\Gamma^\rho_{\mu\nu}$ to the torsionful connection \eqref{eq:NCconnection}. The torsion comes from the fact that for TTNC we have $\tau\wedge d\tau=0$ so that the first term in \eqref{eq:NCconnection} is no longer symmetric. The difference between these two connections is a tensor depending on $b_\mu$. We find
\begin{eqnarray}
&& \mathcal{L}  =  e\left[\left(h^{\alpha\nu}h^{\beta\mu}-h^{\alpha\mu}h^{\beta\nu}\right)K_{\alpha\mu}K_{\beta\nu}+2\hat v^\mu b_\mu h^{\nu\rho}K_{\nu\rho}\right.\nonumber\\
&&\left.-2\left(\hat v^\mu b_\mu\right)^2 -\tilde\Phi\tilde{\mathcal{R}}-2\hat v^\mu f_\mu+2\epsilon^{\mu\nu\rho}\tau_\rho\hat v^\sigma\beta_\nu R_{\mu\sigma}(K)\right]\,.\nonumber\\ \ \label{eq:La}
\end{eqnarray}

If we express the spatial curvature $\tilde{\mathcal{R}}$ in terms of the spatial curvature $\mathcal{R}$ defined with respect to the $\Omega$ connection in \eqref{eq:curlR} we find\footnote{Formula (12.49) of \cite{Hartong:2015zia} contains a typo. The vector $a_\mu$ should have been $b_\mu$. Since $h^{\mu\nu}b_\nu=\frac{1}{2}h^{\mu\nu}a_\nu$ this explains the factor of 2 difference between the expression here and formula (12.49) of \cite{Hartong:2015zia}.}
$\tilde{\mathcal{R}}=\mathcal{R}-\nabla_\mu\left(h^{\mu\nu}a_\nu\right)$. The vector $a_\mu$ is called the acceleration vector in HL gravity. In TTNC geometry it is known as the torsion vector
$
a_\mu=\mathcal{L}_{\hat v}\tau_\mu\,,
$
since all information about the torsion of \eqref{eq:NCconnection} is contained in $a_\mu$. The extrinsic curvatures $K_{\mu\rho}$ obey $h^{\nu\rho}K_{\mu\rho}=-\nabla_\mu\hat v^\nu$. We see that the DeWitt metric has $\lambda=1$ where $\lambda$ is the parameter in HL gravity that measures the relative coefficient of the two extrinsic curvature terms. The difference with \eqref{eq:U(1)invmodel} is that now there are couplings to $\hat v^\mu b_\mu$. We note that $b_\mu$ and $f_\mu$ transform as
$
\delta b_\mu=\partial_\mu\Lambda_D+\Lambda_K\tau_\mu\,,\delta f_\mu=\partial_\mu\Lambda_K+2\Lambda_K b_\mu-2\Lambda_D f_\mu\,,
$
where $\Lambda_D$ and $\Lambda_K$ are the local parameters of the $D$ and $K$ transformations. We can thus gauge fix the $K$ transformations by setting $\hat v^\mu b_\mu$ to any desired value.  

Finally we rewrite the last term in \eqref{eq:La}. Using that for TTNC we can always write
$
\tau_\mu=N\partial_\mu\tau\,,
$
it can be shown that
\begin{equation}\label{eq:LT}
\epsilon^{\mu\nu\rho}\tau_\rho\hat v^\sigma\beta_\nu R_{\mu\sigma}(K)=-\frac{1}{4}\epsilon^{\mu\nu\rho}\beta_\nu\tau_\rho\left(\partial_\mu+2a_\mu\right)I\,,
\end{equation}
where $I$ is defined as
$
I=B^2-4\left(\hat v^\mu b_\mu\right)^2+2\hat v^\nu \partial_\nu\left(B-2\hat v^\mu b_\mu\right)-4\hat v^\mu f_\mu\,,
$
in which $B$ denotes the quantity
$
B=\hat v^\mu N^{-1}\partial_\mu N\,.
$
Our final result is thus \eqref{eq:La} with \eqref{eq:LT}. The action depends on the variables $\tau_\mu=N\partial_\mu\tau$, $\bar h_{\mu\nu}$, $\tilde\Phi$, $\hat v^\mu b_\mu$, $\hat v^\mu f_\mu$ and $\beta_\mu$. The equation of motion for $\beta_\mu$ allows us to solve for $\hat v^\mu f_\mu$ on-shell. 

The Lagrangian \eqref{eq:La} provides a new way of constructing conformal actions for non-projectable HL gravity that we refer to as Schr\"odinger gravity. The main difference with the $z=2$ Weyl invariant construction of \cite{Bergshoeff:2014uea,Hartong:2015zia} is that we do not need to introduce a St\"uckelberg scalar, called $\chi$ in \cite{Bergshoeff:2014uea,Hartong:2015zia}. This St\"uckelberg scalar was needed in order to construct a $z=2$ Weyl invariant combination of extrinsic curvature terms based on a DeWitt metric with $\lambda$ parameter $1/2$, i.e. $\left(h^{\alpha\nu}h^{\beta\mu}-\frac{1}{2}h^{\alpha\mu}h^{\beta\nu}\right)K^\chi_{\alpha\mu}K^\chi_{\beta\nu}$ where $K^\chi_{\mu\nu}$ is the extrinsic curvature scalar with $m_\mu$ replaced by $m_\mu-\partial_\mu\chi$ (see \cite{Hartong:2015zia} for details).

\subsection{Lifshitz solutions}

The Schr\"odinger invariant CS theory \eqref{frlag} with $c_2=c_3=0$ admits $z=2$ Lifshitz solutions. It can be readily verified that the following expressions solve the flatness conditions $F=dA+A\wedge A=0$,
\begin{equation}\label{LifAsol}
\tau=\frac{dt}{r^2}\,,\  e^1=\frac{dr}{r}\,,\  e^2=\frac{dx}{r}\,,\  b=-\frac{dr}{r}\,,\  \beta=-\frac{dx}{r}\,,
\end{equation}
with all other connections equal to zero. If we use the relation to the ADM description of HL gravity expressed in \eqref{eq:ADM} and \eqref{eq:HLmetric} we find the $z=2$ Lifshitz metric
\begin{equation}
ds^2=-\frac{dt^2}{r^4}+\frac{dr^2}{r^2}+\frac{dx^2}{r^2}\,.
\end{equation}
The solution has a simpler form. If we denote $\mathbf{b}= e^{(D-P_1) \rho}$, where $r= e^{-\rho}$, then the Lifshitz solution can be written as $
A= \mathbf{b}^{-1} a  \mathbf{b} + \mathbf{b}^{-1} d\mathbf{b} \,
$,
where $a = H dt + (P_2 -Z) dx$. 

The 3D Lifshitz solution with $z=2$ was also found in the context of CS theories for higher spin theories \cite{Gary:2012ms,Gary:2014mca}. However, it was pointed out in \cite{Lei:2015ika} that this interpretation is problematic due to a degeneracy problem: the spin-connection cannot be determined from the torsion-free equation. Put another way the non-relativistic solutions of $SL(N,\mathbb{R})\times SL(N,\mathbb{R})$ CS theory are not equivalent to metric solutions. Here we show that the solution \eqref{LifAsol} naturally emerges from a Newton--Cartan Chern--Simons theory which is not a Lorentzian metric theory.

\section{Discussion}\label{sec:discussion}

The results obtained in this paper open up for a number of interesting applications and extensions.
First of all, it will be interesting to examine CS actions for other non-relativistic algebras, such as the Galilean conformal algebra,
and likewise for algebras that play a role in ultra-relativistic limits, such as the Carroll algebra. In the latter case, one expects
a connection to the 3D Carrollian gravity of Ref.~\cite{Hartong:2015xda}. 

Another worthwhile direction to pursue is to consider the CS actions of this paper in the presence of non-trivial boundaries, 
and consider aspects of edge physics as performed e.g. in \cite{Gromov:2015fda} for quantum Hall states. In particular it would interesting to study the role of the Galilean boost CS term (with coefficient $c_2$ in \eqref{eq:GboostCS}) in relation to anomalies in this context. Further one could try to find a microscopic description of the extended Bargmann CS theory, e.g. using non-relativistic fermions with a mass gap such that the effective theory below the mass gap is described by the extended Bargmann CS theory\footnote{We thank Kristan Jensen for pointing this out.}.
Moreover it is tempting to consider the CS theory with the Galilean boost and rotation CS terms (with coefficients $c_2$ and $c_3$) in \eqref{eq:CStheory} as the non-relativistic analogue of topologically massive gravity \cite{Deser:1982vy,Li:2008dq}. To explore this idea further one would for example like to understand the solutions of the theory.

An important application of our findings  is to use the Schr\"odinger invariant CS theory as a bulk holographic action
for $z=2$ Lifshitz space-times. The resulting Schr\"odinger gravity may be regarded as a very minimal setup to do Lifshitz holography (see \cite{Taylor:2015glc}
for a review). Using HL gravity in this context was proposed in \cite{Janiszewski:2012nf,Griffin:2012qx} and the CS reformulation of this paper is expected to provide new insights.
In particular the CS formulation can give a proper definition of black objects (provided they exist) in these non-relativistic gravity theories, and 
therewith also give information on boundary hydrodynamics and other dynamical properties. 
We also stress that our results point towards Lifshitz vacua appearing naturally in non-relativistic gravity, rather than in
Lorentzian metric theories. It would thus be interesting to revisit some of the 
pathologies \cite{Copsey:2010ya} and other properties (see e.g. \cite{Keeler:2013msa}) that have been examined within the framework
of Riemannian geometry. 

Another relevant aspect to pursue, in close parallel with higher spin gravity, is to employ
the techniques of  \cite{Ammon:2013hba,deBoer:2013vca} to find the corresponding generalization of holographic entanglement 
\cite{Ryu:2006bv} for non-relativistic CS gravity.  Moreover, a further extension of our ideas to non-relativistic higher spin gravity could be an interesting direction. Similar in spirit, an $SL(2,\mathbb{R}) \times U(1)$ CS theory (called lower spin gravity) was argued to be the minimal setup 
to holographically describe warped CFTs \cite{Hofman:2014loa}. In this light one could try to find a relation between the present CS theories or some close cousin thereof and 2-dimensional warped CFTs \cite{Hofman:2011zj}. 

All the HL gravity actions obtained via our CS formulation have the property that the HL $\lambda$ parameter, which appears in the DeWitt metric contracting the extrinsic curvatures, is equal to unity. It would thus be interesting to see whether by adding appropriate scalar matter fields, i.e.
considering CS matter theories, we can construct more general HL actions for which $\lambda\neq 1$.

Upon the completion of this work we were informed by Eric Bergshoeff and Jan Rosseel of the paper
\cite{Bergshoeff:2016lwr} in which  it is shown that the Bargmann invariant CS action can be obtained by
a non-relativistic limit from three-dimensional GR, augmented with two vector fields. This work also
obtains a supersymmetric generalization, which is thus a supersymmetric extension of 3D projectable HL gravity.  

\section*{Acknowledgments}

We would like to thank  Eric Bergshoeff,  Joaquim Gomis, Daniel Grumiller, Diego Hofman, Kristan Jensen, Wout Merbis and Jan Rosseel for valuable discussions. 
The work of JH is supported by the advanced ERC grant `Symmetries and Dualities in Gravity and M-theory' of Marc Henneaux. The work of NO is supported in part by the Danish National Research Foundation project ``New horizons in particle and condensed matter physics from black holes".  
JH and NO gratefully acknowledge support from the Simons Center for Geometry and Physics, Stony Brook University at which some of the research for this paper was performed.

\bibliography{Lifshitzhydro}
\bibliographystyle{apsrev4-1}
\end{document}